**Title: Characterization of the novel transposon Tn*7722* harboring *bla*$_{\text{NDM-1}}$: Insights into the evolutionary dynamics of resistance in *Klebsiella pneumoniae***

Authors list: Tram Vo[1,2], Aïcha Hamieh[1,2], Marc Levy [3], Pierre Pontarotti [1,2,4], Jean-Marc Rolain [1,2], Vicky Merhej[1,2]

Affiliations:

[1]Microbes, Evolution, Phylogénie et Infection (MEPHI), Assistance Publique-Hôpitaux de Marseille (APHM), Aix Marseille University, 13005 Marseille, France.

[2] Institut Hospitalo-Universitaire (IHU) Méditerranée Infection, 13005 Marseille, France.

[3] Bacteriology-Hygiene Unit, French Polynesia Hospital, Tahiti

[4] Centre National de la Recherche Scientifique (CNRS-SNC5039), 13009 Marseille, France

Correspondence:

Vicky Merhej

Pharm D, PhD

vicky.merhej@univ-amu.fr

Aix Marseille Univ, IRD, APHM, MEPHI, Faculté de Sciences Médicales et Paramédicales, 19-21 boulevard Jean Moulin, 13385 Marseille CEDEX 05, France.

Summary: 250

Text: 4657

References: 52

Figures: 3

Table: 1

Format: original research

**Abstract**

**Background**

*Klebsiella pneumoniae* is a major opportunistic pathogen responsible for various invasive infections. The rise of carbapenem-resistant *K. pneumoniae*, primarily due to acquisition of *bla*$_{\text{NDM}}$ genes, presents a serious global health threat. In French Polynesia, where international travel is frequent, sporadic cases of NDM-producing *Enterobacteriales* have emerged. This study aims to characterize the genomic features of NDM-producing *K. pneumoniae* isolates collected in French Polynesia and evaluate the roles of clonal expansion and horizontal gene transfer mediated by mobile genetic elements in $bla_{NDM}$ spread.

**Materials and Methods**

Between July 2006 and September 2021, 17 carbapenemase-producing *K. pneumoniae* isolates were identified from 715 clinical samples in Tahiti. Whole-genome sequencing using Illumina MiSeq and Oxford Nanopore technologies was performed.

**Results**

Seven NDM-producing *K. pneumoniae* strains were identified, five $bla_{NDM-1}$ and two $bla_{NDM-9}$ variants. All isolates were resistant to ertapenem (MICs 1 to >32 mg/L), with three resistants to imipenem (MICs 8 to >32 mg/L) and six to meropenem (MICs 2 to >8 mg/L). A novel IS*26*-mediated composite transposon, Tn*7722* (16,246 bp), was detected in four isolates on IncF and IncR plasmids. This transposon also carried *qnrS1* and *aph(3')-VI* genes, conferring resistance to fluoroquinolones and aminoglycosides. Tn*7722*-like elements were found in diverse bacterial genomes worldwide, suggesting it facilitates $bla_{NDM}$ transmission across multiple species and regions.

**Conclusion**

NDM-producing *K. pneumoniae* in French Polynesia remain sporadic but genetically diverse, without evidence of local outbreak. It underscores the role of plasmid and Tn*7722*-driven evolution and adaptation. Ongoing genomic surveillance is vital to track the evolution of high-risk clones and MGEs guiding effective containment.

**Keywords**: *Klebsiella pneumoniae*, $bla_{NDM}$, Tn*7722*, carbapenem resistance, mobile genetic elements

## 1. Introduction

*Klebsiella pneumoniae* is a major opportunistic pathogen responsible for severe invasive infections, including pneumonia, intra-abdominal infections, and bloodstream infections, owing to its wide antimicrobial resistance potential [1]. The emergence and rapid global dissemination of carbapenem-resistant *K. pneumoniae* (CRKP) poses a critical threat to public health, as it severely limits effective treatment options and is associated with increased morbidity and mortality [2,3]. This resistance is primarily driven by the acquisition of carbapenemase-encoding genes, mainly $bla_{KPC}$, $bla_{NDM}$ and $bla_{OXA-48}$ which produce enzymes capable of hydrolyzing a broad spectrum of cephalosporins and carbapenems [4]. Among them, the New Delhi metallo-β-lactamase (NDM) is especially concerning due to its extensive hydrolytic activity against most β-lactams, including resistance to many β-lactamase inhibitors, as well as its rapid global dissemination. First reported in 2008 by Yong *et al.,* $bla_{NDM}$ was identified in a *K. pneumoniae* isolate from a Swedish patient who had previously received medical care in India [5]. Since then, 40 distinct NDM variants have been identified and reported globally across multiple bacterial species, including *E. coli*, *Enterobacter* spp., *Pseudomonas*, and *Acinetobacter* [5–7]. Among the NDM variants, $bla_{NDM-1}$ and $bla_{NDM-5}$ are the most prevalent, especially in the Indian subcontinent, China, and Europe, while others like $bla_{NDM-4}$, $bla_{NDM-6}$ and $bla_{NDM-7}$, appear sporadically, indicating ongoing evolution and spread [3,8–10].

The rapid global spread of $bla_{NDM}$ across diverse bacteria is mainly driven by successive horizontal gene transfer, facilitated by mobile genetic elements (MGEs), especially plasmids and transposons [10,11]. Over 20 plasmid incompatibility groups carry $bla_{NDM}$, with IncX3 plasmids being the most prevalent in *Enterobacteriales* due to their high conjugation efficiency and worldwide distribution. Other common vectors include IncFIB, IncFII, IncA/C, IncL/M, and IncHI plasmids; these plasmids are often linked to multidrug resistance. This diversity of plasmid backbones demonstrates $bla_{NDM}$'s adaptability to various genetic contexts. Within plasmids, $bla_{NDM}$ is frequently embedded in complex regions flanked by insertion sequences (ISs) and composite transposons [11,12]. It's spread in *K. pneumoniae* is driven by MGEs like IS*26*, IS*Aba125*, IS*3000* and IS*CR1*, and transposons such as Tn*3*, Tn*125*, Tn*3000*, and Tn*7382* [12–16]. The propagation of $bla_{NDM}$ represents a complex, multilayered process driven by diverse MGEs that enable its horizontal transfer via recombination, transposition, and plasmid-mediated conjugation [10,17].

Extrinsic anthropogenic factors, including international travel and global trade, significantly accelerate the global spread of $bla_{NDM}$. Numerous studies have highlighted international travel as a key route for introducing $bla_{NDM}$ into previously unaffected regions. Travelers often acquire NDM-producing *Enterobacteriales* during their stay and can remain asymptomatically colonized for extended periods after returning home, with colonization frequently persisting for several months and strongly correlated with the travel destination. A compelling example of the complex dissemination dynamics of $bla_{NDM}$ is provided by Oueslati et al. 2021, who documented the polyclonal spread of NDM-1 and NDM-9-producing *Escherichia coli* and *K. pneumoniae* in Tahiti, French Polynesia, since 2015 [18]. Although French Polynesia's status as a major international tourist destination may facilitate repeated introductions of $bla_{NDM}$ through travel, only one patient in this study had a documented history of international travel [18]. This suggests that transmission likely occurred locally, either within the community or via environmental reservoirs in healthcare settings. The identification of diverse plasmid incompatibility groups, including IncR, IncHI2, and IncF, supports the role of horizontal gene transfer in disseminating $bla_{NDM}$ among genetically unrelated strains. While international travel serves as a primary route for the introduction of $bla_{NDM,}$ into new regions; the sustained persistence and global dissemination of this resistance gene are largely driven by local environmental and healthcare reservoirs. Under continuous antibiotic selection pressure, MGEs promote the maintenance and horizontal transfer of $bla_{NDM,}$ within microbial communities, facilitating its long-term spread and shifting transmission from epidemic outbreaks to endemic persistence [18].

In French Polynesia, plasmid-mediated horizontal transfer drives $bla_{NDM}$ spread among *E. coli* and *K. pneumoniae*, but the role of transposons remains unclear. To address this gap, we performed a longitudinal genomic analysis of NDM-producing *K. pneumoniae* clinical isolates collected over a 16-year period in Tahiti, with a focus on the genetic environment surrounding $bla_{NDM}$ to better understand its long-term dissemination and evolutionary dynamics. Our results reveal that the involvement of transposons in $bla_{NDM}$ mobilization and rearrangement is even more extensive than previously anticipated, highlighting their critical contribution to the gene's persistence and adaptability within local bacterial populations.

## 2. Materials and Methods

### 2.1. Bacterial strains

Between 2006 and 2021, a total of 715 non-duplicate *K. pneumoniae* clinical isolates were collected at the *centre hospitalier de la Polynésie Française* (Tahiti hospital). As part of a continuous infection control and antimicrobial resistance surveillance program, all hospitalized patients underwent weekly rectal swab screening to detect colonization with extended-spectrum β-lactamase (ESBL)-producing and carbapenemase-producing *Enterobacteriales*. The screening policy was applied uniformly throughout the study period. Isolates were defined as CRKP based on clinical microbiology laboratory testing performed at the hospital using routine antimicrobial susceptibility testing in accordance with EUCAST or CLSI guidelines. From this collection, 17

*K. pneumoniae* isolates exhibiting resistance to at least one carbapenem (ertapenem, imipenem, or meropenem) were selected for further investigation. These CRKP isolates were subsequently sent to the IHU-Méditérranée Infection research laboratory in Marseille, France, for detailed analysis of their resistance mechanisms.

## 2.2. Species identification and antimicrobial susceptibility testing

Bacterial species were identified using matrix-assisted laser desorption ionization–time of flight mass spectrometry (MALDI-TOF MS) (Bruker Daltonics, Bremen, Germany) [19]. Antimicrobial susceptibility testing (AST) was performed using the disk diffusion method on Mueller-Hinton agar with a panel of 16 antibiotics (SirScan Discs; i2a, France). Results were interpreted according to EUCAST 2023 guidelines (https://www.eucast.org/). Minimum inhibitory concentrations (MICs) for ertapenem, imipenem and meropenem were determined by Etest (BioMérieux, Marcy l'Étoile, France). The MIC of colistin was determined by broth microdilution (Biocentric, Bandol, France). *K. pneumoniae* ATCC 13883 was used as the quality control strain.

## 2.3. Carbapenemase detection and molecular characterization

Carbapenemase activity was assessed using the Carba NP test (Biorad, Hercules, California, United States), according to the original protocol by Nordmann *et al.* [20].This colorimetric assay detects imipenem hydrolysis by carbapenemase-producing strains, indicated by a pH-dependent color shift of phenol red from red to yellow.

Genomic DNA was extracted from CRKP isolates using the EZ1 BioRobot system with the "Bacterial DNA" protocol and kit (Qiagen, Hilden, Germany). DNA from CRKP isolates was extracted by the automated EZ1 method (Qiagen BioRobot EZ1-, Tokyo, Japan) using "Bacterial DNA" program and kit (EZ1 DNA, Qiagen, Hilden, Germany), according to the manufacturer's instructions [21]. Detection of carbapenemase genes was performed by standard or quantitative PCR (qPCR) targeting class A, B and D carbapenemase genes including $bla_{\text{KPC}}$, $bla_{\text{NDM}}$, $bla_{\text{VIM}}$, $bla_{\text{OXA-48}}$, $bla_{\text{OXA-23}}$, $bla_{\text{OXA-24}}$, and $bla_{\text{OXA-58}}$ using specific primers and probes, as previously described [22].

## 2.4. Whole-genome sequencing (WGS) and bioinformatic analysis of CRKP isolates.

Genomic DNA of NDM-producing *K. pneumoniae* isolates was extracted using an automated EZ instrument (Qiagen, Hilden, Germany) following the manufacturer's protocol, ensuring high purity and yield suitable for downstream sequencing applications. Whole-genome sequencing (WGS) was performed using Illumina MiSeq technology (Illumina Inc., San Diego, CA, USA) with a paired-end read configuration ($2 \times 150$ bp), allowing for high accuracy in base calling and coverage uniformity. Raw sequencing reads underwent quality control using FastQC v0.12.1 [23]. Subsequently, reads were trimmed and filtered using Trimmomatic v0.36 [24] to remove low-quality bases and adapter sequences, improving assembly accuracy. Additionally, long-read sequencing was conducted using Oxford Nanopore Technology (ONT; Oxford, United Kingdom). Hybrid genome assemblies were generated with Unicycler v0.5.0 [25], integrating both short- and long-read data to achieve complete chromosomal and plasmid sequences of carbapenem-resistant *K. pneumoniae* isolates. Assembly quality was evaluated using QUAST v5.2.0 [26] and genome annotation was performed with Prokka v1.14.6 [27]. Multilocus sequence typing (MLST) was conducted using MLST v2.22.0 via the Galaxy Australia platform (https://usegalaxy.org.au/), enabling classification of isolates into sequence types (STs) and facilitating epidemiological comparisons.

## 2.5. Phylogenomic analysis of NDM-producing *K. pneumoniae* genomes

To investigate the evolutionary relationships among NDM-producing *K. pneumoniae* strains, we performed a comparative phylogenomic analysis incorporating both the genomes analyzed in this study and publicly available reference genomes harboring $bla_{\text{NDM}}$ genes. The pangenome analysis was conducted using Roary v3.13.0 employing a BLASTp identity threshold of 95% and defining core genes as those present in 99% of the genomes analyzed. The resulting core genome alignment was used as the input for phylogenetic inference. A maximum likelihood phylogenetic construction was generated with RaXML software v8.2.12 using default parameters. The phylogenetic tree was visualized and annotated using the Interactive Tree of Life (iTOL) platform (https://itol.embl.de/).

### 2.6.        Identification of resistance genes and MGEs

To investigate the genetic determinants of antimicrobial resistance and their association with mobile genetic elements (MGEs), a comprehensive *in silico* analysis was conducted. Antibiotic resistance genes (ARGs) were identified using Abricate v1.0.1, employing the Resfinder [28] and Comprehensive Antibiotic Resistance Database CARD [29] as reference sources. Genes were considered present if they met thresholds of at least 90% sequence identity and 70% coverage. Plasmid content was assessed using two complementary tools: PlasmidFinder, for replicon typing and the mob-typer module of MOB-suite v3.0.3, which identifies relaxase genes (MOB types), assigns plasmid mobility classes (conjugative, mobilizable, or non-mobilizable), and provides plasmid clustering based on sequence similarity [30]. Conjugative elements were identified using oriTfinder, which detects origin-of-transfer (oriT) sites and conjugation-associated genes, including components of the type IV secretion system. IntegronFinder v2.0.2 was used to detect integrons by identifying integrase genes and associated recombination sites [31]. Insertion sequences (ISs) were annotated using the ISfinder database [32]. To investigate transposable elements involved in resistance gene mobility, genome assemblies were screened against the Tncentral database [33] and the Transposon registry ([https://transposon.lstmed.ac.uk/tn-registry](https://transposon.lstmed.ac.uk/tn-registry)), enabling classification and contextual analysis of known transposons implicated in antimicrobial resistance dissemination.

## 3. Results

### 3.1.     Characterization of carbapenem-resistant isolates

Seventeen carbapenem-resistant out of 715 (2.38%) isolates initially identified as *K. pneumoniae* resistant to at least one carbapenem antibiotic, were received from Tahiti Hospital, collected between 2006 and 2021 from various clinical specimens, across different hospital units. Species identification using MALDI-TOF MS confirmed that all isolates (100%) were *K. pneumoniae*. Phenotypic confirmation of carbapenemase production was obtained using the Carba NP test, with seven *K. pneumoniae* isolates testing positive. These NDM-producing isolates were identified between 2014 and 2020. They were mainly recovered from rectal swabs (n = 4), indicating colonization, and urine samples (n = 3), reflecting clinical infection (Table 1). Molecular characterization by PCR confirmed the presence of $bla_{NDM}$ carbapenemase genes in all isolates, with five harboring the $bla_{NDM-1}$ variant and two carrying $bla_{NDM-9}$. No other carbapenemase genes, including the $bla_{KPC}$, $bla_{OXA-48}$ or $bla_{VIM}$ were identified in the seventeen isolates highlighting the predominance of NDM enzymes within this collection. Co-occurrence of resistance genes was common; all isolates also carried extended-spectrum β-lactamase (ESBL) genes and various combinations of $bla_{SHV}$, $bla_{TEM}$ and $bla_{CTX-M-15}$ genes were observed (Fig. 1). Phenotypic susceptibility testing demonstrated that all isolates were resistant to *β*-lactams, including amoxicillin/clavulanic acid, cefepime, piperacillin/tazobactam, ceftriaxone, and ceftazidime (Table 1). Minimum inhibitory concentration (MIC) testing for carbapenems showed elevated resistance levels in all isolates. Ertapenem resistance was observed in all seven strains, with MICs ranging from 1.5 to over 32 mg/L. Imipenem and meropenem MICs varied widely, from 0.38 to over 32 mg/L and 1 to over 8 mg/L, respectively. Six of the seven isolates were resistant to at least two carbapenems, confirming the high-level resistance commonly associated with NDM production (Table 1). Importantly, all isolates remained susceptible to aztreonam and colistin, with MICs below 1 mg/L (Table 1).

The NDM-producing *K. pneumoniae* isolates displayed a multidrug-resistant (MDR) phenotype, with resistance encompassing both β-lactam and non-β-lactam antibiotic classes (Table 1). All isolates carried genes encoding aminoglycoside-modifying enzymes (AACs, ANTs, APHs), five harbored *qnrS1* and two carried *qnrB* genes conferring resistance to Fluoroquinolone. All isolates possessed *fosA* gene conferring fosfomycin resistance (Fig. 1). These findings underscore the heterogeneous resistance profiles among the NDM-producing *K. pneumoniae* isolates and point to the coexistence of multiple resistance mechanisms, likely acquired via horizontal gene transfer.

### 3.2. Polyclonal lineages and plasmid-mediated *bla*$_{NDM}$ dissemination

To investigate the genomic characteristics and epidemiological relationships of NDM-producing *K. pneumoniae* isolates, whole-genome sequencing (WGS) was performed on the seventeen clinical isolates collected between 2014 and 2020. Sequencing yielded high-quality assemblies, with an average genome size of approximately 5.7 Mb, N50 values ranging from 4.58 to 5.48 Mb, and mean coverage depths exceeding 112-fold (Table S1). These metrics ensured the reliability of downstream analyses, including multilocus sequence typing (MLST) and resistance gene profiling (Fig. 1, Table S2). The genome assemblies and associated metadata have been made publicly available through the GenBank database under Bioproject accession number PRJNA1201927, supporting further comparative and epidemiological studies (Table S1).

All seven NDM-producing isolates were assigned to distinct sequence types (STs), specifically ST22, ST34, ST111, ST147, ST307, ST661, and ST1967, underscoring the substantial genetic diversity within the collection (Fig. 1, Table S2). The absence of shared STs among the isolates indicates a lack of clonal dissemination and suggests multiple independent introductions or emergences of $bla_{NDM}$ carbapenemase genes in the region (Fig. 1). Comprehensive plasmid profiling of the seven *K. pneumoniae* isolates revealed a high degree of diversity in plasmid content, indicating the involvement of multiple plasmid backbones in the dissemination of carbapenemase genes (Fig. 2). Using PlasmidFinder, each isolate was found to harbor between two and six distinct plasmid replicons, encompassing more than ten incompatibility (Inc) groups. The most prevalent replicon types included IncFIB, IncFII, and IncR, followed by IncHI1B and IncX3 (Table S3). All seven isolates harbored $bla_{NDM}$ genes located on plasmids, including two variants $bla_{NDM-1}$ and $bla_{NDM-9}$ (Fig. 1). A total of 30 plasmids were identified, of which 15 were predicted to be conjugative, three mobilizable and 12 non-mobilizable (Table S3). Both $bla_{NDM-1}$ and $bla_{NDM-9}$ genes were located on plasmids predicted to be conjugative or mobilizable, facilitating their dissemination among bacterial populations (Fig. 2).

The $bla_{NDM-9}$ gene was detected on nearly identical 265 kb conjugative IncH-type plasmids, designated pQ8532-1 and pQ8534-1, which were carried by *K. pneumoniae* isolates Q8532 (ST34) and Q8534 (ST307), respectively. These plasmids harbored a complete *tra* operon encoding both DNA-processing proteins, including relaxase, and mating pair formation (MPF) components enabling autonomous conjugative transfer (Fig. 2). Comparative genomic analysis revealed that pQ8532-1 and pQ8534-1 shared extensive sequence identity with a previously described IncH12 plasmid isolated from *K. pneumoniae* ST2570 (CP031850) in China in 2018, yet this reference plasmid does not harbor the $bla_{NDM-9}$ gene (Fig. S1). In addition to $bla_{NDM-9}$, these plasmids carry a diverse array of antibiotic resistance determinants, including two class 1 integrons (*intI1*) harboring gene cassettes conferring resistance to aminoglycosides, sulfonamides, and fosfomycin, thereby functioning as multidrug resistance platforms. The presence of the *mer* operon and *qacE* gene further confers resistance to mercury and antiseptic compounds, respectively, emphasizing the plasmids' adaptive advantage in environments subjected to multiple selective pressures (Fig. 2, Table S3).

The $bla_{NDM-1}$ gene was detected in five *K. pneumoniae* isolates Q8539, Q8535, pQ8529, Q8533 and Q8540, carried on three distinct plasmid types. They included the 63.4 kb IncR plasmid pQ8539-3, the 163 kb IncA/C2 conjugative plasmid pQ8535-1 and the highly identical 126 kb IncFIB/IncFII plasmids pQ8529-1, pQ8533-2 and pQ8540-2 (Fig. 1, 2). Comparison with publicly available plasmid sequences in NCBI revealed that plasmid pQ8535-1 showed high similarity with a *Salmonella enterica* plasmid (KR091911.1) which contains $bla_{NDM-1}$ (Fig. S1). The plasmid encoded an *intI1* carrying resistance genes to aminoglycosides, trimethoprim, chloramphenicol, β-lactams ($bla_{CMY-59}$), and antiseptics (*qacE*) (Fig. 2). Additionally, it harbored a complete *tra* operon, facilitating autonomous conjugative transfer. The plasmid pQ8539-3 showed high similarity to an IncR plasmid carrying $bla_{NDM-1}$ from a clinical *K. pneumoniae* strain isolated in Thailand in 2016 (LC613144) (Fig. S1). The plasmid carried an *intI1* with additional resistance genes including $bla_{TEM-12}$ and *qnrS1*, conferring resistance to β-lactams and fluoroquinolones, respectively (Fig. 2). Although pQ8539-3 encoded a relaxase capable of initiating DNA processing at the origin of transfer (oriT), the absence of MPF components rendered it non-conjugative (Fig. 2). The plasmids pQ8529-1, pQ8533-2 and pQ8540-2 were closely related to pK66-45-3 (CP020904.1), a ~120 kb IncFIB/IncFII plasmid from *K. pneumoniae* strain K66-45 (GCF-002113865.1), isolated in Norway in 2010 [34,35] (Fig. S1). They encoded the MPF conjugation system with a *tra* operon and carried multiple $\beta$-lactamase genes, including $bla_{\text{TEM-150}}$, $bla_{\text{OXA-9}}$ and $bla_{\text{CTX-M-15}}$ (Fig. 1, 2). Interestingly, while pK66-45-3 itself lacked $bla_{\text{NDM-1}}$, it coexisted with a second 338.5 kb conjugative InH/IncF/IncR plasmid, pK66-45-1 (CP020902.1), which harbored $bla_{\text{NDM-1}}$ within a region flanked by structural features characteristic of transposable elements. This genetic environment warrants further investigation of its potential role in the mobilization of $bla_{\text{NDM-1}}$ (Fig. S1, 3A).

### 3.3. Characterization of Tn*7722*: A novel composite transposon haboring $bla_{\text{NDM-1}}$

In-depth analysis of the genomic context surrounding $bla_{\text{NDM-1}}$ in the plasmids pQ8529-1, pQ8533-2 and pQ8540-2 revealed its localization within a contiguous 16,246 bp element flanked by two directly oriented IS26 copies generating 5'-GAAAAT-3' target site duplications, consistent with a composite transposon structure (Fig. 3A). Its configuration is distinct from previously described $bla_{\text{NDM-1}}$-associated transposons such as Tn*3000* and does not correspond to any previously registered element from the Transposon Registry (https://transposon.lstmed.ac.uk/tn-registry) (Fig. S2, Table S4). Add to this, it showed no significant homology to any known transposable elements in ISFinder or NCBI databases, both at the nucleotide and protein levels. BLASTn analysis of Tn*7722* against five well-characterized $bla_{\text{NDM-1}}$-carrying transposons (Tn*125*, Tn*3000*, Tn*6935*, Tn*6960*, Tn*6923*) revealed high nucleotide homology exceeding more than 99% (Fig. S2). However, the coverage of Tn*7722* by these elements was partial, ranging from 22% for Tn*3000* to 65% for Tn*6360* (Table S2). This indicates that while Tn*7722* shares highly conserved regions particularly around *bla*$_{NDM-1}$ with known transposons, it represents a distinct overall genetic structure. Altogether, these features define a previously undescribed transposon, for which the name Tn*7722* has been assigned (https://transposon.lstmed.ac.uk/tn-registry?page=101).

Tn*7722* encodes a complete DDE-type transposase (Tn*2*, 826 amino acids) followed by a resolvase-like recombinase gene (*tnpR*), a complex array of mobile elements including IS*Ecp36,* IS*Spu2*, IS*Kpn19* and IS*Ab125* alongside *qnrS1* and *aph(3')-VIa* genes conferring resistance to fluoroquinolones and aminoglycosides, respectively (Fig. 3A). Compared with Tn*3000*, Tn*7722* retains the conserved core segment found in *bla*$_{NDM-1}$-carrying transposons, namely, IS*Aba125*-*bla*$_{NDM-1}$-*ble*$_{MBL}$-*trpF-dsbD-cutA*, but it lacks both flanking IS3000 elements and the *grosES* and *groEL* at the 3′ end due to an IS*26* insertion [12,36] (Fig. 3A). Additionally, Tn*7722* incorporates a complete Tn*6292* element carrying *qnrS1* and *aph(3')-VIa*. These features indicate that Tn*7722* has undergone multiple recombination and rearrangement events, resulting in a mosaic structure distinct from the reference Tn*3000* (Fig. 3A, S2) The organization of Tn*7722* is as follows: IS*26*-Tn*2*-IS*Ecp36*-*qrnS1*-*tnpR*-IS*Kpn19*-*aph(3')-VIa*-IS*Aba125*-IS*Spu2*-*bla*$_{NDM-1}$-*ble*$_{MBL}$-*trpF-dsbD-cutA*-IS*26*. Such structural diversification of composite transposons may enhance the horizontal transfer and co-selection of multidrug resistance genes among diverse bacterial hosts and plasmids.

Comparative genomic analysis showed that plasmids pQ8529-1, pQ8533-2, and pQ8540-2 share a highly conserved backbone with the conjugative plasmid pK66-45-3 from *K. pneumoniae* ST11 strain K66-45 (Fig. 3A). Although pK66-45-3 lacks Tn*7722*, a second plasmid from the same strain, pK66-45-1, harbors the complete Tn*7722*, which represents the only region shared with the other plasmids. The conserved backbone architecture among these plasmids indicates a homologous relationship, suggestive of a shared evolutionary origin. The coexistence of pK66-45-1 and pK66-45-3 within the same host, together with the identification of target site duplications flanking Tn7722, supports a cut-and-paste mechanism of intra-cellular transposition (Fig. 3A,B) This likely facilitated the transfer of $bla_{NDM1}$ between co-resident plasmids, underscoring the role of Tn*7722* as a MGE accelerating antimicrobial resistance spread. Similarly, in isolate Q8539, $bla_{NDM1}$ was located on a non-conjugative IncR plasmid (pQ8539-3), while the same strain also carried a conjugative plasmid (pQ8539-1) sharing 98.54% identity with a Tn*7722*-positive plasmid from *K. pneumoniae* ST395 (Table S5). These observations reinforce Tn*7722*'s capacity for intra-host mobilization, promoting gene exchange across diverse plasmid types. A search of the PLSDB database (https://ccb-microbe.cs.uni-saarland.de/plsdb/) identified Tn*7722*-like elements (≥ 99% nucleotide identity) on 57 plasmids, predominantly from *K. pneumoniae* (n=54), but also from *K. variicola*, *E. coli* and unclassified *Enterobacteriaceae* strain (Table S5). These plasmids were affiliated with diverse incompatibility groups, including IncF, IncH, IncR, and IncU and were isolated from various continents and countries in the world, including Russia, USA, Italy, Germany, Korea, China… (Table S5). Tn*7722* was frequently flanked by truncated IS*26* elements, often disrupted at the 5′ end due to the insertion of additional MGEs carrying a diverse array of ARGs, including IS*Ecp1*-$bla_{CTX-M-15}$, $bla_{TEM-12}$, $bla_{OXA-18}$, *ant1*, *aacA4*, among others (Fig. 3A, S3). These structural rearrangements reflect the ongoing evolutionary plasticity of Tn*7722* and its role in accumulating and disseminating multidrug resistance determinants.

## 4. Discussion

This study presents comprehensive, longitudinal genomic surveillance of carbapenem-resistant *K. pneumoniae* isolates collected over a 16-year period in hospitals in Tahiti. Consistent with previous reports by Oueslati et al., molecular typing and genomic analysis confirmed the polyclonal nature of $bla_{NDM}$-positive *K. pneumoniae* in this region, identifying seven isolates assigned to distinct sequence types (ST22, ST34, ST111, ST147, ST307, ST661, and ST1967) [18]. The marked genetic diversity and absence of repeated sequence types indicate multiple independent introductions rather than local clonal expansion. These findings are in line with observations from other geographically isolated, low-endemic settings such as Reunion Island, where sporadic cases of NDM-producing *K. pneumoniae* have been linked primarily to travel or medical repatriation rather than sustained transmission [37]. By contrast, regions with high endemicity often experience large-scale clonal outbreaks driven by epidemic lineages [38,39]. For example, the nationwide dissemination of $bla_{NDM-1}$ ST11 strains in Poland [40] and prolonged outbreaks of ST147 in Southern Italy [41] illustrate how such clones can rapidly dominate once established. The presence of globally distributed high-risk clones, including ST147 and ST307, among the Tahitian isolates underscores the ongoing risk posed by international movement of multidrug-resistant organisms [42]. Together, these results highlight the critical need for continuous genomic surveillance in non-endemic, remote healthcare settings. Early detection of sporadic introductions allows for timely infection control interventions, helping to prevent the establishment and spread of high-risk *K. pneumoniae* clones and protecting vulnerable healthcare systems from emerging multidrug resistance threats.

Our findings underscore the central role of plasmids in the dissemination of $bla_{NDM}$, particularly those that are conjugative and mobilizable, which serve as efficient vectors for horizontal gene transfer [43]. The localization of $bla_{NDM}$ on such plasmids facilitates its rapid spread not only within individual species but also across species and even genus boundaries, significantly contributing to the global propagation of carbapenem resistance [11]. These plasmids frequently harbor multiple resistance determinants, conferring broad-spectrum antimicrobial resistance through co-selection, and are commonly associated with high-risk *K. pneumoniae* clones such as ST147 and ST307 [42]. The global rise in healthcare-associated infections caused by these clones is largely attributable to their remarkable capacity to acquire a wide range of plasmids harboring carbapenemase genes, including $bla_{NDM}$, across diverse incompatibility groups such as IncFIIA, IncA/C, and IncX3. ST147 and ST307 not only maintain these plasmids stably but also act as donors, facilitating interspecies and intergenus transfer of $bla_{NDM}$ to *Escherichia coli*, *Enterobacter* spp., and other *K. pneumoniae* clones, thereby fueling successive nosocomial outbreaks [43]. Importantly, these plasmids often carry additional genetic elements that enhance bacterial fitness in hospital settings. In particular, resistance genes against antiseptics that are frequently co-located with antibiotic resistance determinants on IncF plasmids suggesting a selective advantage in environments with high antimicrobial and biocidal pressure. This dual capacity to confer resistance to both antibiotics and antiseptics likely contributes to enhanced persistence, environmental survival, and host colonization. Taken together, the convergence of high-risk clonal lineages with highly transmissible, multiresistant, and environmentally adaptive plasmids represents a critical threat to infection control. These findings highlight the need for integrated genomic surveillance strategies that simultaneously track both clonal dissemination and plasmid dynamics as key drivers in the epidemiology of carbapenem resistance.

We report the discovery of Tn*7722*, a novel composite transposon carrying $bla_{NDM}$, embedded within diverse plasmid backbones. Tn*7722* displays hallmark features of IS*26*-mediated transposition, including formation of circular intermediates, which enable its mobilization beyond the confines of a single plasmid [12–15,44]. This unique "Russian doll" genetic architecture allows Tn*7722* to overcome plasmid-host specificity barriers by facilitating the horizontal transfer of *bla*$_{NDM}$ between plasmids, thereby driving the dissemination of carbapenem resistance across a wide range of bacterial hosts and plasmid types [45]. Such transposon-mediated mobility is particularly significant in clinical environments where multiple plasmid types coexist and interact, fostering extensive horizontal gene transfer and contributing to the rapid emergence and spread of multidrug-resistant pathogens [12,13,46]. This dynamic mechanism highlights the critical role of composite transposons like Tn*7722* in shaping the epidemiology of resistance by promoting the exchange of resistance determinants beyond traditional plasmid boundaries. Moreover, Tn*7722* encodes multiple resistance determinants alongside *bla*$_{NDM}$, promoting co-selection and the consolidation of multidrug resistance. This co-localization complicates treatment by conferring resistance to multiple antibiotic classes and underscores the adaptive advantage of composite transposons under strong antimicrobial pressure. Beyond gene dissemination, transposons such as Tn*7722* are central drivers of plasmid evolution and plasticity. Through genomic rearrangements and acquisition of novel resistance genes, these mobile elements facilitate dynamic remodeling of plasmid architecture, enhancing bacterial fitness and persistence in clinical environments [10–13,44,47–49]. The mosaic structure of these transposons allows plasmids to selectively acquire resistance determinants as needed, minimizing replication burden while maximizing adaptability. This genomic flexibility accelerates plasmid diversification, thereby complicating infection control efforts and highlighting the necessity for surveillance strategies that encompass gene transfer mediated by transposons within plasmids.

The highly similar genetic architecture of Tn*7722*-carrying plasmids from this study and publicly available sequences suggests a shared evolutionary origin, most plausibly explained by intra-cellular transposition of Tn*7722* between co-resident plasmids within the same host, followed by structural rearrangements of the element (Fig. 3). The presence of a complete Tn*7722* element on plasmid pK66-45-1 supports its role as the probable donor. In contrast, derivative plasmids show evidence of truncation at the 5′ end of Tn*7722*, marked by an incomplete IS*26* element and often accompanied by additional mobile elements such as IS*Ecp1*-*bla*$_{CTX-M-15}$, indicating post-transposition modifications (Fig. 3) [50]. The presence of Tn*7722* on mobilizable and conjugative plasmids like pQ8529, pQ8533, and pQ8540 highlights its role in broadening the host range of resistance genes across *Enterobacteriaceae*. This mechanism illustrates the dynamic interplay between transposons and plasmids, driving the rapid evolution and horizontal spread of multidrug resistance under the selective pressures of clinical settings (Fig. 3B). The close relatedness of plasmids from this study to pK66-45-3, isolated from *K. pneumoniae* strain K66-45 in Norway from a patient previously hospitalized in India, highlights the geographic and temporal context of Tn*7722* transposition during a period when *bla*$_{NDM-1}$ was becoming endemic in India [5,6,35,51]. This underscores the central role of international travel and medical tourism in driving the global dissemination of resistance elements, such as plasmids carrying Tn*7722* with *bla*$_{NDM-1}$ , and illustrates how human movement shapes the global epidemiology of antimicrobial resistance. In regions such as French Polynesia, human mobility, including travel and short-term stays, may facilitate the intermittent introduction of *K. pneumoniae* strains harboring plasmids with Tn*7722* carrying *bla*$_{NDM-1}$, enhancing the spread of resistance genes across geographic regions (Table S2). The mobility of Tn*7722*-bearing plasmids further promotes the efficient dissemination of *bla*$_{NDM-1}$ across diverse bacterial hosts and plasmid backbones. This synergy between human movement, plasmid conjugation, and transposon activity accelerates the local and global spread of multidrug resistance. Additionally, terrestrial livestock production systems and aquatic environments in such regions may act as reservoirs for resistance genes, underscoring the need for a One Health approach to monitor and mitigate the spread of antimicrobial resistance across human, animal, and environmental domains [52].

In conclusion, our study reveals the complex genomic landscape of NDM-producing *K. pneumoniae* in French Polynesia and identifies the novel transposon Tn*7722* as a pivotal driver of $bla_{\text{NDM-1}}$ dissemination. Tn*7722*'s capacity to mobilize across diverse plasmid backbones accelerates the global dissemination of carbapenem resistance among various bacterial species, a process intensified by human travel and population movement. Our findings underscore the critical interplay between mobile genetic elements and human movement in shaping antimicrobial resistance patterns. They also highlight the persistent challenge posed by diverse resistance reservoirs entering local settings, emphasizing the urgent need for comprehensive molecular surveillance and stringent infection control measures, especially in regions with high tourism, to effectively monitor and curb the spread of NDM-producing pathogens.

**Declarations**

**Funding:** This work was supported by the French Government under the "Investissements d'avenir" (Investments for the Future) programme managed by the Agence Nationale de la Recherche (ANR, fr: National Agency for Research), (reference: Méditerranée Infection 10-IAHU-03).

**Competing Interests:** The authors declare that they have no competing interests.

**Ethical approval**

Ethical approval was not required for this study, as all isolates were obtained during routine clinical laboratory procedures and analyzed anonymously. No patient identifiers were collected, and there was no direct patient involvement or intervention.

## Acknowledgements

The authors are very grateful to the IHU Méditerranée-Infection for financial support. They sincerely thank Gael Duclaux for strain retrieval, identification, and shipment, and Laurent Dortet for his support and collaboration.

## GenBank Accession numbers

The genome sequences of NDM-producing *K. pneumoniae* isolates were deposited in Genbank database under Bioproject accession number PRJNA1201927, ensuring data accessibility for further research and validation (Table S1). Genbank accession of isolates, Q8529 (CP177086-CP177088), Q8532 (JBKFEZ000000000), Q8533 (CP177089-CP177091), Q8534 (CP177052-CP177057), Q8535 (CP176849-CP176855), Q8539 (JBKOUA000000000), Q8540 (CP1177092-CP1177097). **References**

## List of table and figures

**Table**

**Table 1.** Characterizations of NDM-producing *K. pneumoniae* isolates from French Polynesia.

**Figure**

**Figure 1.** Phylogenetic analysis and resistome profiling of NDM-producing *Klebsiella pneumoniae* isolates from French Polynesia, 2014–2021.The heatmap shows the number of antibiotic resistance genes (ARGs) detected per isolate. ARGs: antibiotic resistance genes; ST: sequence type; DCs: drug classes.

**Figure 2.** Plasmids encoding *bla*$_{NDM}$ genes, other resistance determinants associated with various MGEs, found in NDM-producing *Klebsiella pneumoniae* isolates from French Polynesia, described in this study.

**Figure 3.** Structure of composite transposon Tn*7722* and schematic representation of its transfer along with *bla*$_{NDM-1}$ and other resistance determinants across various bacterial species.

**Supplementary data**

**Table S1.** Genomes assembly of NDM-producing *Klebsiella pneumoniae* isolates from French Polynesia genomes.

**Table S2.** Sequence types (STs) and allelic profiles of NDM-producing *Klebsiella pneumoniae* isolates from French Polynesia genomes.

**Table S3.** Genomic characterisation and MGEs in NDM-producing *Klebsiella pneumoniae* isolates from French Polynesia.

**Table S4.** Blastn analysis of Tn*7722* with reference transposons carrying *bla*$_{NDM-1}$.

**Table S5.** Informations of Tn*7722*-carrying plasmids in different continents.

**Figure S1.** Sequence alignment of *bla*$_{NDM-9}$-carrying plasmids (A), *bla*$_{NDM-1}$-carrying plasmids (B) from *Klebsiella pneumoniae* isolates in French Polynesia, described in this study, and reference plasmids from the NCBI database. Alignment plots were generated using the EasyFig tool.

**Figure S2.** Comparison of Tn*7722* with other transposons in the Transposon Registry database.

**Figure S3.** Comparison of Tn*7722* transposons of plasmids detected in different continents.

**Table 1. Characterizations of NDM-producing *K. pneumoniae* isolates from French Polynesia.**

| Characterizations | NDM-producing *K. pneumoniae* isolates | | | | | | |
|---|---|---|---|---|---|---|---|
| | Q8529 | Q8533 | Q8535 | Q8539 | Q8540 | Q8532 | Q8534 |
| Source | Urine | Rectal swab | Rectal swab | Urine | Rectal swab | Urine | Rectal swab |
| Year of Isolation | 2014 | 2018 | 2019 | 2020 | 2021 | 2017 | 2018 |
| Hospital Unit | Emergency | Hematology | Hematology | Intensive care | Hematology | Emergency | Cardiological Intensive care |
| NDM variants | NDM-1 | NDM-1 | NDM-1 | NDM-1 | NDM-1 | NDM-9 | NDM-9 |
| ESBL[a] | SHV-187 TEM-1 CTX-M-15 | SHV-27 TEM-150 CTX-M-15 | SHV-26 TEM-35 | SHV-187 TEM-166 | SHV-11 TEM-1A CTX-M-15 | SHV-106 TEM-206 CTX-M-15 | SHV-26 |
| AmpC | - | - | CMY-2 | - | - | - | DHA-1 |
| **MIC (mg/L)[a]** | | | | | | | |
| Ertapenem | >32* | 4* | 8* | 1.5* | 1.5* | 2* | 1* |
| Imipenem | >12* | 1.5 | >32* | 8* | 1 | 1.5 | 0.38 |
| Meropenem | >8* | 4* | >8* | 2* | 2* | >8* | 1 |
| Colistin | 0.25 | 0.25 | 0.25 | 0.25 | 1 | 0.5 | 0.125 |
| **Antimicrobial susceptibility testing[b]** | | | | | | | |
| Amoxicillin | R | R | R | R | R | R | R |
| Amoxicillin + CLA | R | R | R | R | R | R | R |
| Cefepime | R | R | R | R | R | R | R |
| Piperacillin + TZB | R | R | R | R | R | R | R |
| Ceftriaxone | R | R | R | R | R | R | R |
| Ceftazidime | R | R | R | R | R | R | R |
| Aztreonam | S | S | S | S | S | S | S |
| Gentamycin | S | S | R | R | R | R | R |
| Amikacin | R | R | R | S | R | R | S |
| Ciprofloxacin | S | R | S | R | R | R | R |
| SXT | S | R | R | R | S | S | S |
| Fosfomycin | R | R | R | R | R | R | R |
| Doxycycline | S | S | S | S | R | R | S |

[a] MIC of ertapenem > 0.5 mg/L, imipenem > 4 mg/L, resistant, meropenem >2 mg/ L, resistant, colistin > 2 mg/L, *, resistant
[b] CLA, clavulanic acid; TZB, tazobactam; SXT, trimethoprim/sulfamethoxazole; R, resistant; S, sensible

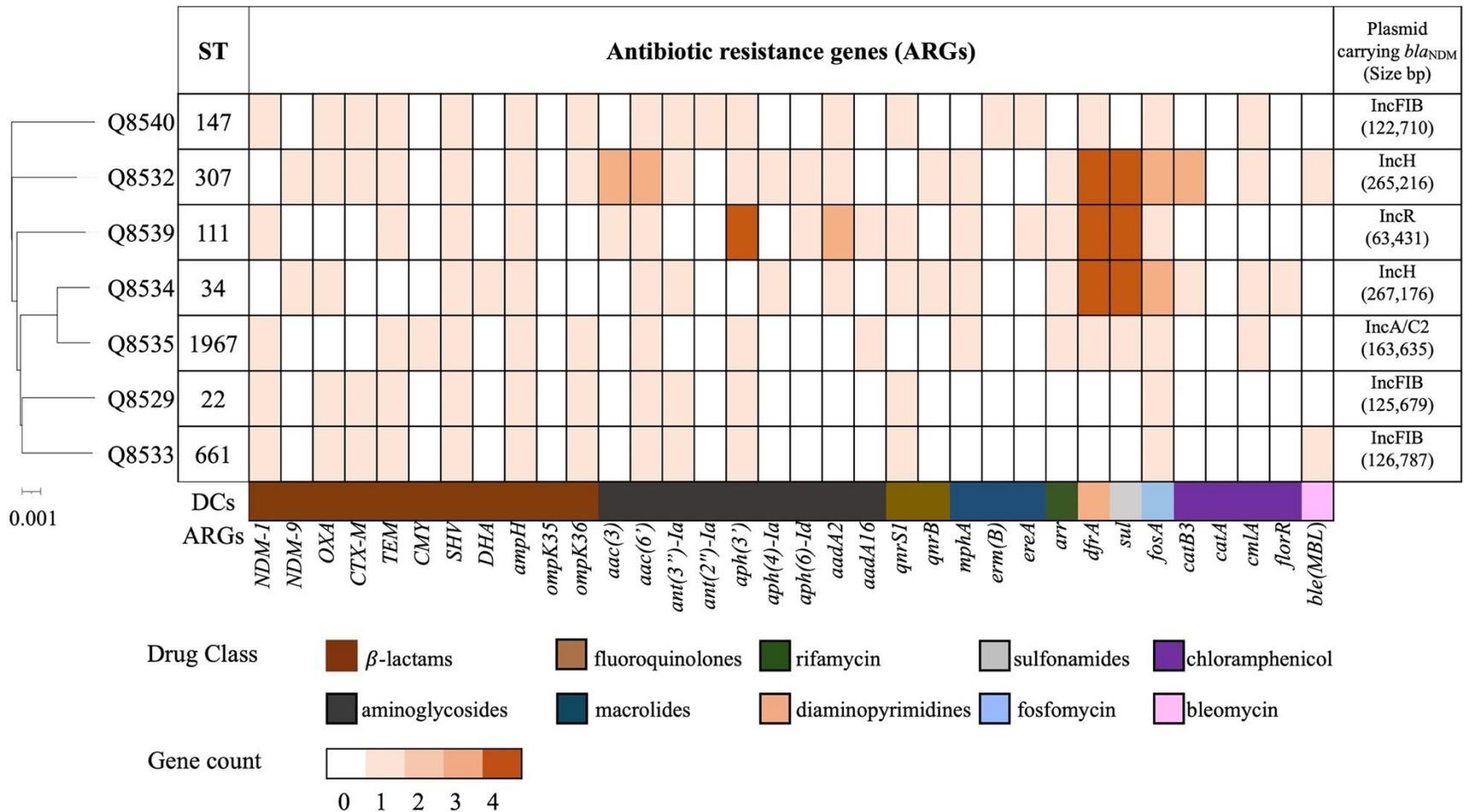

**Figure 1.** Phylogenetic analysis and resistome profiling of NDM-producing *Klebsiella pneumoniae* isolates from French Polynesia.

The heatmap shows the number of antibiotic resistance genes (ARGs) detected per isolate. ARGs: antibiotic resistance genes; ST: sequence type; DCs: drug classes.

**Figure 2.** Plasmids encoding *bla*NDM genes, other ARGs associated with various MGEs, found in NDM-producing *Klebsiella pneumoniae* isolates from French Polynesia, described in this study.

**Figure 3.** Structure of composite transposon Tn*7722* and schematic representation of its transfer along with *bla*$_{NDM-1}$ and other resistance determinants across various bacterial species.

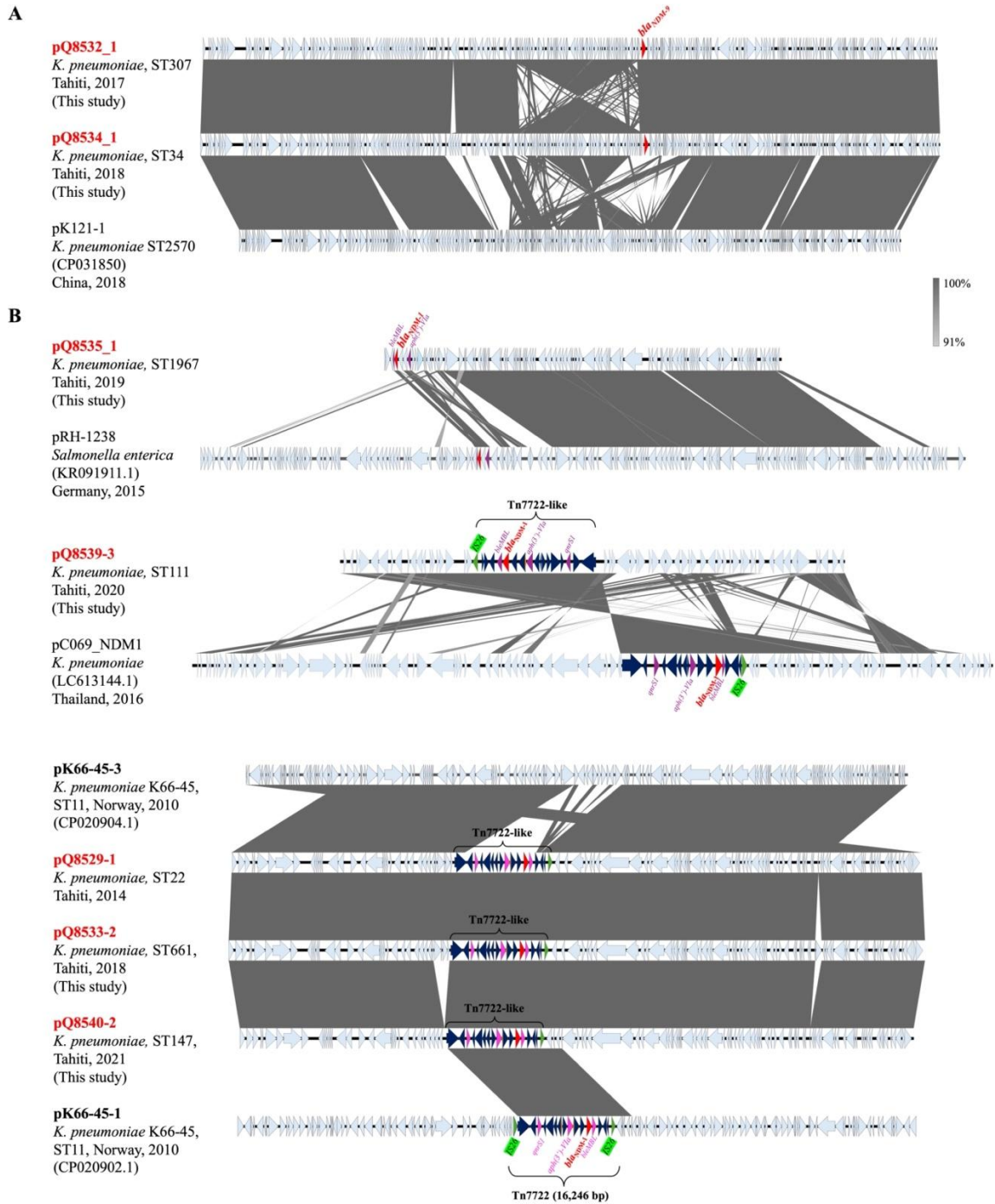

**Figure S1.** Sequence alignment of *bla*NDM-9-carrying plasmids (A), *bla*NDM-1-carrying plasmids (B) from *Klebsiella pneumoniae* isolates in French Polynesia, described in this study, and reference plasmids from the NCBI database. Alignment plots were generated using the EasyFig tool.

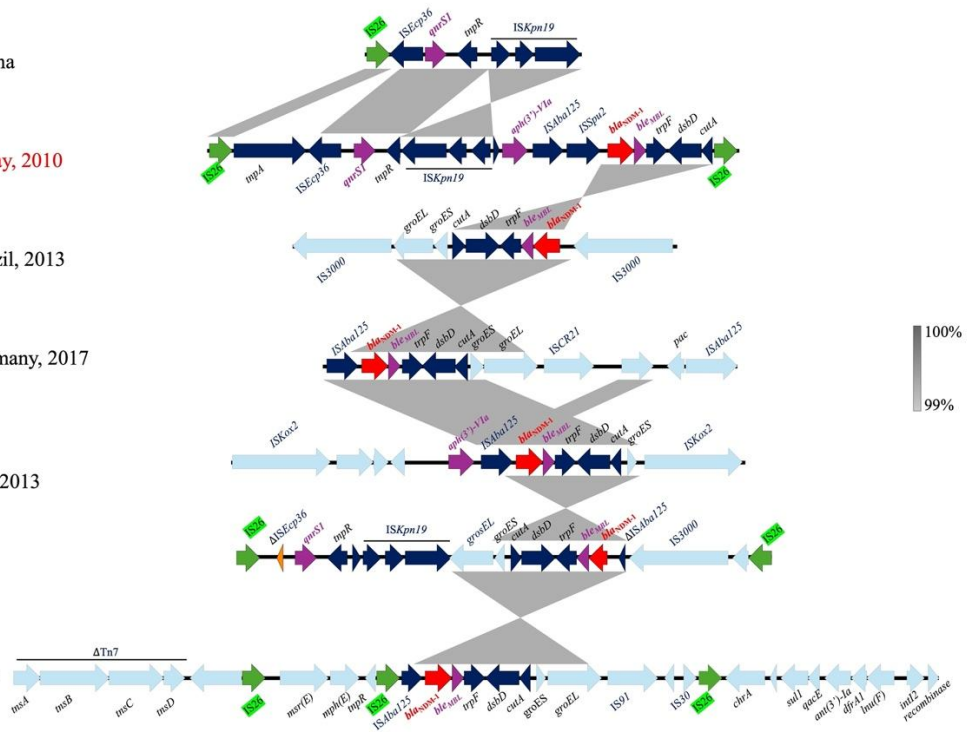

**Figure S2.** Comparison of Tn*7722* with other transposons in the Transposon Registry database.

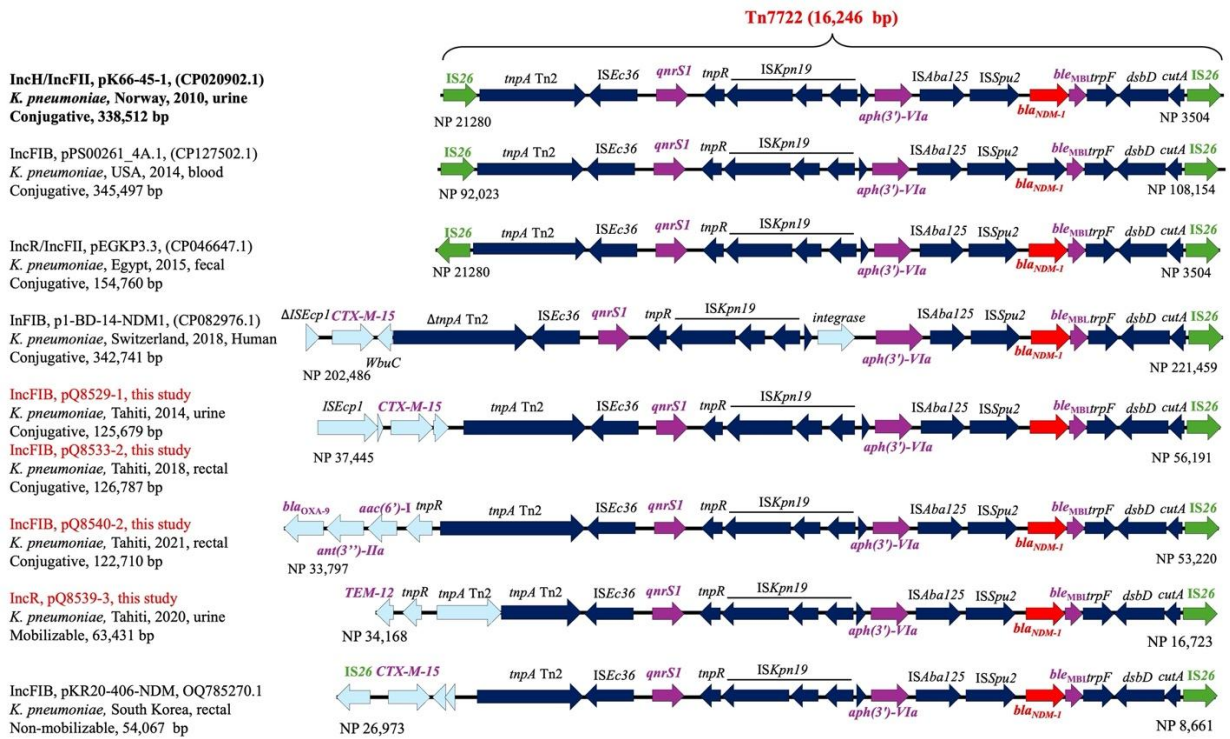

**Figure S3.** Comparison of Tn*7722* transposons of plasmids detected in different continents.